# DAΦNE INTERACTION REGIONS UPGRADE

C. Milardi, INFN/LNF, Frascati (Roma), Italy for DAΦNE Collaboration Team[1].

*Abstract*

DAΦNE, the Frascati Φ-factory, has recently completed experimental runs for the three main detectors, KLOE, FINUDA and DEAR achieving $1.6 \times 10^{32}$ cm$^{-2}$s$^{-1}$ peak and 10 pb$^{-1}$ daily integrated luminosities.

Improving these results by a significant factor requires changing the collision scheme. For this reason, in view of the SIDDHARTA detector installation, relevant modifications of the machine have been realized, aimed at implementing a new collision scheme based on a large Piwinski angle and *crab-waist*, together with several other hardware modifications involving injection kickers, bellows and beam pipe sections.

## INTRODUCTION

DAΦNE [2] is a lepton collider, working at the c.m. energy of the Φ resonance (1.02 GeV). It provided high K meson rates to three different experiments: KLOE, DEAR and FINUDA, taking data one at a time.

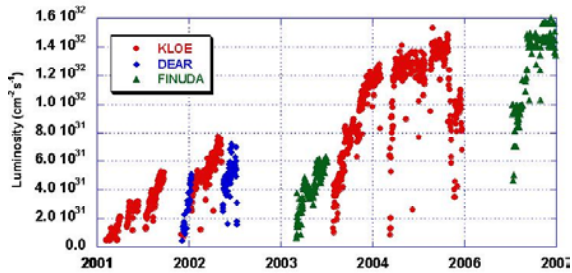

Figure 1: DAΦNE daily peak luminosity trend.

In its original configuration the collider consisted of two independent rings, each ~ 97m long, sharing two interaction regions IR1 and IR2 where the KLOE [3] and DEAR [4] or FINUDA [5] detectors were respectively installed. A full energy injection system, including a S-band linac, 180 m long transfer lines and an accumulator/damping ring, provides the e$^+$ and e$^-$ beams with the required emittance and energy spread.

The DAΦNE complex runs also a beam test facility, providing e$^-$/ e$^+$ beams from the linac in the energy range 25÷725 MeV with tunable intensity from $10^{10}$ to a single particle per pulse.

A synchrotron radiation facility with three independent beam lines, collecting the radiation emitted in one wiggler and two bending magnets of the e$^-$ ring, is also available.

In these years DAΦNE has undergone several progressive upgrades [6,7,8], aimed at improving the collider performances (see Fig.1), implemented during the shut-downs for detectors changeover.

## ESTABLISHED PERFORMANCES

Since 2001 the DAΦNE accelerator complex has been delivering luminosity to three experiments, improving, at the same time, its performances in terms of luminosity, lifetime and backgrounds.

The DEAR experiment has been done in less than 5 months during 2002-2003, collecting about 200 pb$^{-1}$, with a peak luminosity of $0.7 \times 10^{32}$ cm$^{-2}$s$^{-1}$.

The KLOE experimental program has been completed in 2006, more than 2 fb$^{-1}$ have been acquired on the peak of the Φ resonance, while more than 0.25 fb$^{-1}$ have been stored off-resonance [7], to perform a high statistics resonance scan. The best peak luminosity obtained has been $1.5 \times 10^{32}$ cm$^{-2}$s$^{-1}$, with a maximum daily integrated luminosity about 10 pb$^{-1}$.

The second run of FINUDA, which collected 1.2 fb$^{-1}$, started in April 2006. During the operation a peak luminosity of $1.6 \times 10^{32}$ cm$^{-2}$s$^{-1}$ has been achieved, while a maximum daily integrated luminosity similar to best during the KLOE run has been obtained with lower beam currents, lower number of bunches and higher beta functions at the collision point [8].

However, these performances were the best obtainable with the DAΦNE original collision scheme.

Long-range beam-beam interactions (parasitic crossings) [9, 10] lead to a substantial lifetime reduction of both beams in collision, limiting the maximum storable current and, as a consequence, the achievable peak and integrated luminosity.

The minimum value of $\beta^*_y$ at the IP is set by the longitudinal bunch size to avoid destructive effects coming from the *hourglass* effect. The bunch length in the DAΦNE main rings is presently, after a careful coupling impedance optimization [11], 25 mm for both beams at the operating bunch current (~ 15 mA). Moreover the horizontal crossing angle at the IP must be lower with the Piwinski limit of 30 mrad.

A new conceptual approach is necessary to push the luminosity towards $10^{33}$ cm$^{-2}$s$^{-1}$. After long studies and discussions involving the Accelerator Division Team and the international accelerator community, a new collision scheme based on large Piwinski angle and *crab-waist* has been adopted for the DAΦNE collider.

## NEW COLLISION SCHEME

The new collision regime [12] devised for DAΦNE is based on a large Piwinski angle Φ obtained by increasing the collision angle θ and reducing the transverse horizontal beam size σ$_x$.

$$\Phi \approx \frac{\sigma_z}{\sigma_x}\frac{\theta}{2}$$

Luminosity and tune-shift depend on the number N of particles in the colliding bunches, on the transverse beam sizes σ$_{x,y}$ and the vertical betatron function $\beta^*_y$ according to the following formulas [13]:

$$L \propto \frac{N\xi_y}{\beta_y^*}; \quad \xi_y \propto \frac{N\sqrt{\beta_y^*}}{\sigma_z \theta}; \quad \xi_x \propto \frac{N}{(\sigma_z \theta)^2}$$

A first luminosity gain can be obtained, while keeping the vertical tune shift constant, by increasing N proportionally to ($\sigma_z\theta$).

Moreover the length of the overlap region of the colliding bunches drops as:

$$\Sigma \propto \frac{\sigma_x}{\theta}$$

Being $\Sigma \ll \sigma_z$, $\beta_y^*$ can be made as small as $\Sigma$ and further advantages can be envisaged in terms of higher luminosity, vertical tune shift reduction and synchrobetatron resonance reduction. However collider operation with large Piwinski angle requires a compensation mechanism for the new beam-beam resonances introduced by the configuration itself and limiting the maximum achievable vertical tune-shift value.

Table 1: DAΦNE beam parameters.

|  | DAΦNE (KLOE) | DAΦNE (Upgrade) |
|---|---|---|
| $I_{bunch}$ (mA) | 13 | 13 |
| $N_{bunch}$ | 110 | 110 |
| $\beta^*_y$ (mm) | 17 | 6 |
| $\beta^*_x$ (mm) | $1.7 *10^3$ | $0.2 *10^3$ |
| $\sigma^*_y$ (μm) | 5.4 | 2.6 |
| $\sigma^*_x$ (μm) | 0.7 | 0.2 |
| $\sigma^*_z$ (mm) | 25. | 20. |
| $\Theta_{cross}/2$ (mrad) | 12.5 | 25. |
| $\Phi_{Piwinski}$ | 0.36 | 2.5 |
| $L$ (cm$^{-2}$s$^{-1}$) * $10^{32}$ | 1.5 measured | 10. expected |

This compensation is provided by a couple of sextupoles installed in symmetric positions with respect to the IP, in phase with it in the horizontal plane and at π/2 in the vertical one: they are called the *crab-waist* sestupoles. They mainly suppress the betatron and sinchrobetatron resonances coming from the vertical motion modulation due to the horizontal oscillation.

The beam parameters adopted to implement the new collision regime at DAΦNE are listed in Table 1. According to the theoretical simulations the new collision scheme should provide a peak luminosity of the order of $10^{33}$ cm$^{-2}$ s$^{-1}$.

## INTERACTION REGIONS EVOLUTION

Large collision angle, crab-waist and small $\beta^*_y$ require important changes in the design criteria of the mechanical and magnetic layout of IR1 [14], see Fig. 3.

The second interaction region has been also completely rebuilt in order to provide full beam separation and in order to be ready, with minor modifications, for a future FINUDA run based on the new collision scheme.

Four beam position monitors, installed both halves of the IR1 and of the ring-crossing region, after the pipe separation, provide beam independent closed orbit measurement even in collision.

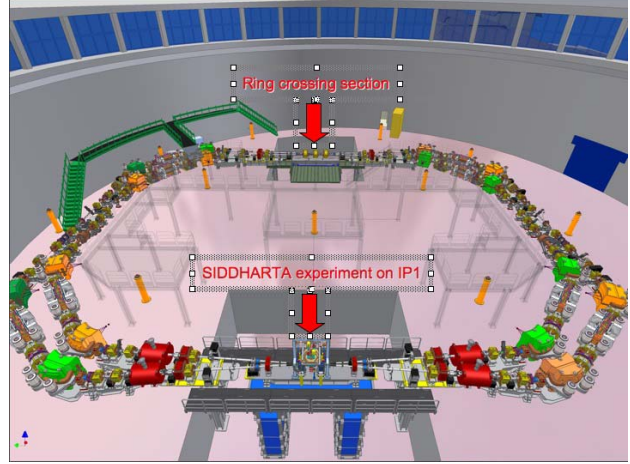

Figure 2: New Main Rings layout

### Interaction region for the SIDDHARTA experiment

Removing the splitter magnets and rotating the two sector dipoles in the long and short arcs adjacent to the interaction regions of both rings has doubled the horizontal crossing angle in IR1.

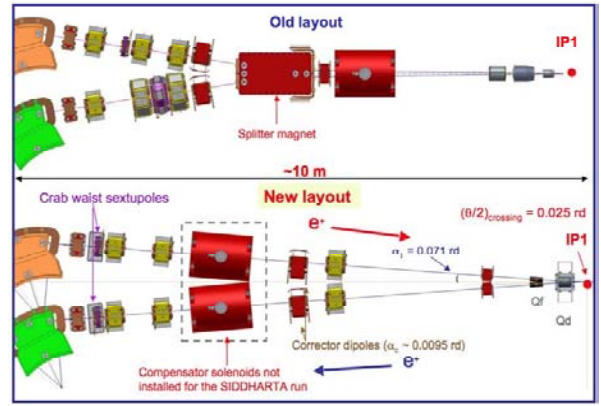

Figure 3: Half view of old IR1 (top) and new (bottom) layout.

The bending fields have been changed, according the values reported in Table 2, in order to meet the new layout angles.

Table 2: Bending dipole parameters; old values are shown in parenthesis.

|  | α [rd] | ρ [m] | B [T] |
|---|---|---|---|
| Sector long | 0.7874 (0.8639) | 1.53 (1.40) | 1.11 |
| Sector short | 0.7834 (0.7069) | 1.27 (1.40) | 1.34 |

Four additional corrector dipoles have been used to match the vacuum chamber in the arcs [15].

The *low-beta* section in the SIDDHARTA IR is based on permanent magnet quadrupole doublets. The quadrupoles are made of SmCo alloy and provide gradients of 29.2 T/m and 12.6 T/m for the first one from the IP and the second one respectively. The first is horizontally defocusing and is shared by the two beams; due to the off-axis beam trajectory, it provides strong beam separation. The second quadrupole, the focusing one, is installed just after the beam pipe separation and is therefore on axis see Fig. 4. The new configuration almost cancels the problems related to beam-beam long range interactions, because the two beams experience only one parasitic crossing inside the defocusing quadrupole where, due to the large horizontal crossing angle, they are very well separated ($\Delta x \sim 20\ \sigma_x$). It is worth reminding that in the old configuration the colliding beams had 24 parasitic crossing in the IRs and in the main one the separation at the first crossing was $\Delta x \sim 7\ \sigma_x$ [9].

The crab-waist sextupoles are installed at both ends of the interaction region. They are electromagnetic devices and the required integrated gradient $k_s$ [16] is:

$$k_s = \frac{1}{2\theta} \frac{1}{\beta_y^* \beta_y^{sext}} \sqrt{\frac{\beta_x^*}{\beta_x^{sext}}}$$

In the case of the optics for the SIDDHARTA operation $k_s = 36.74\ m^{-2}$, more than a factor 5 larger than the average required for the normal sextupoles used for chromaticity correction.

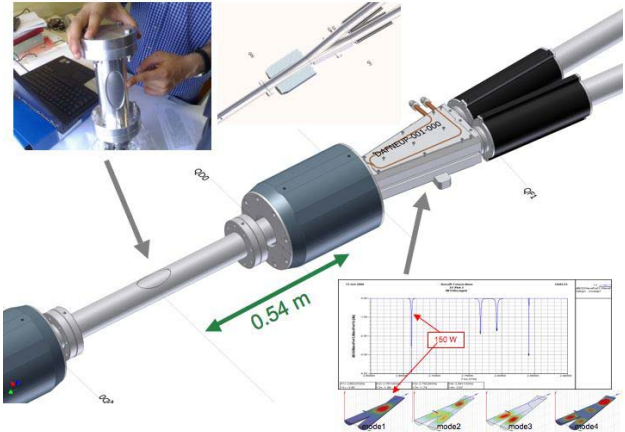

Figure 4: IR1 low-beta section (center), detail of the thin window at IP1 (upper left), vacuum chamber cross section view (upper right) and HOM analysis in the *y-section* (bottom right).

Four electromagnetic quadrupoles have been installed on both sides of IP1 to get the proper phase advance between the *crab-waist* sestupoles and the interaction point.

The compensator solenoids, present in the original setup, have been removed since there is no solenoid around the SIDDHARTA detector. However there is room to reintroduce them for a possible future KLOE run. In a such case, due to the new geometrical layout of IR1, two compensator solenoids will be necessary for each ring, requiring an upgrade of the cryogenic transfer lines.

The DAΦNE main rings layout evolution is shown in Fig. 2.

### Ring crossing region

A new crossing section providing complete separation between the two beams has replaced the second interaction region [15]. It is geometrically symmetric to IR1 and its vacuum chamber is based on the same design criteria. Independent beam vacuum chambers are obtained by splitting the original pipe in two *half-moon* shaped sections, see Fig. 5, providing full vertical beam separation

This aspect is quite relevant because it cancels completely the problems coming from the beam-beam long range interaction [10], allowing at the same time to relax the ring optics requirements imposed by beam separation at the unused interaction point.

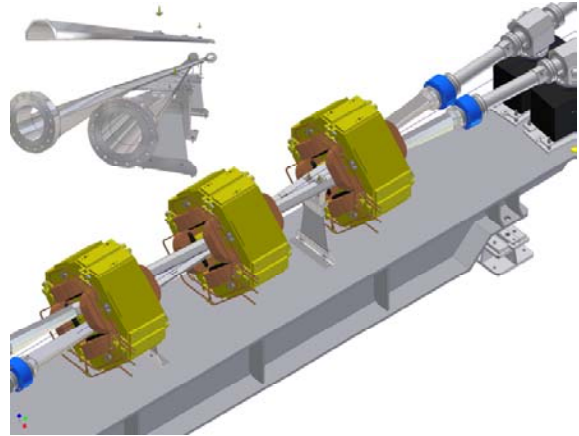

Figure 5: Quadrupole triplet in the ring crossing region (bottom) and *half-moon* vacuum chamber design (top).

The magnetic layout of the ring crossing region is the same as in IR1, but for the missing *crab-waist* sextupoles. and the central focusing section. It is based on a single large aperture electromagnetic quadrupoles triplet, allowing for wide operation flexibility in terms of betatron functions.

### IR Vacuum chamber

The design criteria of the new vacuum chamber for IR1 and the ring crossing region are very simple. All the possible discontinuities have been avoided in order to keep the ring coupling impedance low. The number of bellows has been also limited to the strict necessary to compensate thermal strain and mechanical misalignments; there are four bellows per ring both in IR1 and in the second crossing region.

The vacuum chamber of IR1 consists of straight pipes merging in a *Y* shaped section. The pipe is aluminum (AL6082) made and is equipped on IP1 with a thin 0.3 mm window.

Special attention has been paid to the *Y-section* design since beam induced electromagnetic fields can generate trapped high order modes (HOM). Simulations have pointed out four possible HOMs, among them only the first is trapped and even in the worst case, when the beam spectrum is in full coupling with the mode, the released power is less than 200 W. Nevertheless the *Y- section* has been equipped with a cooling system to remove the heating due to the HOM [17].

*Bellows*

New bellows have been developed and installed in the new IR1 and in the ring crossing section. Presently four new bellows are used in each one of the previous sections.

They connect circular cross section pipes of 88 mm diameter. The inner radius of bellows convolutions is ≈ 65 mm, the outer one 80 mm and the length ≈50 mm, see Fig. 6. Their innovative component is the RF shield [17], necessary to avoid the discontinuity acting as a cavity for the beam. The new RF shield is implemented by means of Ω shaped Be-Cu strips, installed all around two cylindrical aluminum shells fixed at the bellows ends.

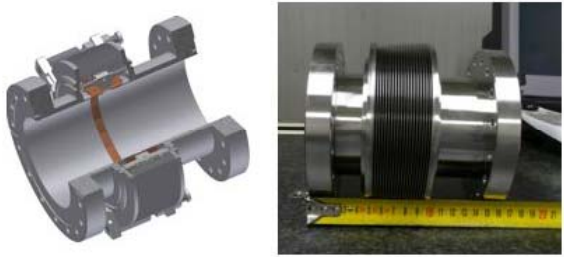

Figure 6: Copper-Beryllium strip shielded bellows, mechanical design (left) and real device (right).

The shield in the old design was realized by using contiguous mini bellows. Experience with the old devices has shown that they were loosing elasticity while aging; moreover they might no longer provide shield contour uniformity when compressed.

HFSS simulations in the frequency range from DC to 5 GHz have shown that the new design reduces bellows contribution to the ring coupling impedance.

## OTHER UPGRADES

*New fast injection kickers*

The injection kickers, two in each Main Ring, have been replaced with new devices [18] based on tapered strips embedded in a rectangular cross section vacuum chamber allowing injection rate up to 50 Hz. The deflecting field is provided by the magnetic and the electric fields of a TEM wave traveling in the structure, which generates 5.4 ns flat top pulses, perturbing only three bunches out of 110 usually colliding. This new injection scheme represents a relevant improvement with respect to the old one, which was based on injection kickers having 150 ns pulse length perturbing almost half of the bunch train.

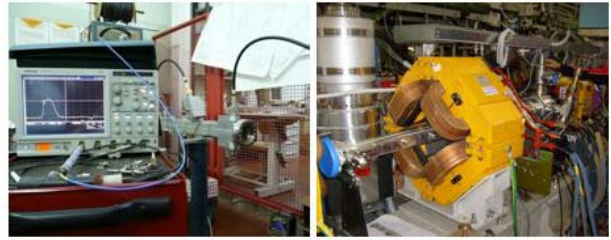

Figure 7: New fast injection kicker under test (left) and installed on the electron ring (right).

Moreover a smooth beam pipe and tapered transitions reduce the kickers contribution to the total ring coupling impedance. All these features should improve the maximum storable currents, colliding beams stability and background on the experimental detector during injection.

*Control System*

A new commercial processor (Pentium/Linux) has been implemented in the control system; this will progressively replace the original home designed front-end processor now seventeen years old.

*Removed and repositioned elements*

Few ion clearing electrodes still installed on the electron ring and no longer necessary have been removed.

The transverse horizontal position of two wigglers in the long arcs has been moved (-2.5 mm) for both rings in order to reduce the non-linear term in the magnetic field predicted by simulations and affecting the beam dynamics.

Positions of the electromagnetic quadrupoles, in the long straight sections on both rings, have been changed to allow the installation of the new injection kickers and to provide a flexible configuration for tuning the phase advance between the two injection kickers themselves.

*RF cavity working frequency*

The new ring layout is ~10 cm shorter than the original one due to the removal of the splitter magnets and the requirement to keep the position of the arcs unchanged in order to minimize the implementation work. As a consequence, the frequency of the RF cavities has been changed by ~ 400 KHz. The variation is well within the tune range of the main rings cavities, but imposes some modifications on the damping ring operating conditions. In fact its RF cavity operates on a sub-multiple frequency of the Main Rings one and the energy variation has to be corrected by changing the dipole field. The tuner range of the cavity has been also adapted in order to be compatible with the new operating conditions.

The front-end electronics used to acquire signals from the beam position monitors required also some modifications in order to work properly with the new frequency value.

*Luminosity monitors*

The new luminosity monitor for the collider consists of three different devices: a small angle Bhabha tile calorimeter split into 20 sectors (30 degrees each) made of alternating lead and scintillating tiles, covering a vertical acceptance between 17.5 and 27 degrees; a GEM (Gas Electron Multiplier) tracker placed in front of the tile calorimeters allowing a redundant measurement of Bhabha events to minimize background; two Single Bremsstrahlung gamma detectors [19].

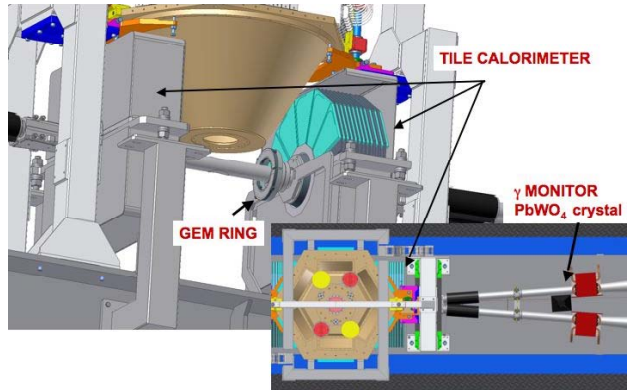

Figure 8: Three-dimensional (top) and top (bottom) view of the SIDDHARTA detector and the luminosity monitors installed around the IP1.

Redundancy in the luminosity measurement is required by the need to best quantify the luminosity gain obtained by adopting the new collision approach. In fact it is of interest not only for DAΦNE but also for the other lepton colliders and even for the LHC hadron collider expected to come in operation soon at CERN.

## CONCLUSIONS

In a five months shutdown the DAΦNE collider has been upgraded to implement a new collision scheme based on large Piwinski angle and *crab-waist*. Commissioning started at the end of November 2007.

Presently the two beams have been stored, all the diagnostic and the new systems have been debugged and put in operation, the new injection kickers and the bellows behave well as espected .

Measurements on the bunch length show a 15% percent reduction at 10 mA per bunch, in agreement with the lower ring impedance.

Preliminary tests with the beams in collision have been also done and look quite promising.

Detailed results and measurements will be published as soon as possible after careful analysis.